# Phonon-induced magnetization dynamics in Co-doped iron garnets

A. Frej[1,a)], C.S. Davies[2], A. Kirilyuk[2], and A. Stupakiewicz[1]

[1]Faculty of Physics, University of Bialystok, Ciolkowskiego 1L, 15-245 Bialystok, Poland
[2]FELIX Laboratory, Radboud University, Toernooiveld 7, 6525 ED Nijmegen, The Netherlands

**Abstract.** The developing field of strain-induced magnetization dynamics offers a promising path toward efficiently controlling spins and phase transitions. Understanding the underlying mechanisms is crucial in finding the optimal parameters supporting the phononic switching of magnetization. Here, we present an experimental and numerical study of time-resolved magnetization dynamics driven by the resonant excitation of an optical phonon mode in iron garnets. Upon pumping the latter with an infrared pulse obtained from a free-electron laser, we observe spatially-varying magnetization precession, with its phase depending on the direction of an external magnetic field. Our micromagnetic simulations effectively describe the magnetization precession and switching in terms of laser-induced changes in the crystal's magneto-elastic energy.

The excitation of the crystal lattice with laser irradiation in the infrared spectral range opens a plethora of possibilities for achieving control over magnetic order. The resonant excitation of high-frequency optical phonon modes can provide a direct and selective ability to manipulate magnetization. It has been shown, for example, that such driving of phonons can result in an effective magnetic field and the creation of a magnon.[1] The wavelength dependence of laser-induced demagnetization of an iron garnet by intense THz pulses proves the importance of frequency-matching.[2] Similarly, ultrafast coherent magnetic phase transitions in $DyFeO_3$ can be driven by the resonant excitation of phonons.[3] Laser-induced strain can briefly change the crystal structure and thus modify the magnetization,[4,5] as exemplified by the recently-reported possibility of using mid-infrared optical pulses to manipulate the domain structure of the antiferromagnet nickel oxide.[6] It is also possible for spin waves to hybridize with acoustic

---

a) Author to whom correspondence should be addressed: a.frej@uwb.edu.pl

waves, forming magnon-polarons.[7,8] Constantly emerging new materials, which allow one to control spins and magnetic ordering through the activation of the crystal's lattice vibrations, has prompted intensive research to focus on thoroughly studying light-induced phonon-magnon interactions.

Recently, we demonstrated the resonant ultrafast phononic switching of magnetization in Co-doped yttrium iron garnet (YIG:Co) using pulses of light within the spectral range 20-40 THz.[9] The rich spectrum of phonon modes characteristic of yttrium iron garnets in this frequency range[10,11] offers a broad platform for such investigations. Studies of the excitation and decay times of these phonon modes provide valuable guidance in identifying which phonon modes could effectively support the manipulation of magnetization.[12] While the phononic switching of magnetization clearly depends on the use of optical pulses with frequencies matching the characteristic frequencies of longitudinal optical phonons, the resulting switching was only measured in Ref. 9 across equilibrium timescales, using magneto-optical microscopy to record the final state of the addressed magnetization. Tracking the magnetization precession actually involved in the switching process would provide insight into the phonon-magnetic reversal mechanisms and possibly reveal the speed of the switching process.

In this Letter, we present experimental measurements and micromagnetic numerical simulations studying time-resolved magnetization dynamics driven by the excitation of optical phonons. Our pump-probe measurements, utilizing single terahertz optical pulses delivered by a free-electron laser, allow us to experimentally track the magnetization precession within an iron garnet thin film. Complimentary micromagnetic simulations are used to predict and understand the observed effects. Our results directly reveal the ultrafast magnetization dynamics that can be triggered by driving optical phonons at resonance.

The experimental studies of the phonon-induced magnetization dynamics were performed using a 7.5-μm-thick iron garnet thin film with composition $Y_2CaFe_{3.9}Co_{0.1}GeO_{12}$. The sample was grown on a gadolinium gallium garnet (GGG) substrate. At room temperature, the constants of cubic and uniaxial anisotropy are $K_1 \approx$ -1 kJ/m$^3$ and $K_u \approx$ -0.1 kJ/m$^3$, corresponding to anisotropy fields of about 290 mT and 28 mT, respectively. The four easy magnetization axes are close to <111>-type orientations.[13,14] The saturation magnetization of YIG:Co at room temperature was 7 kA/m, and the Gilbert damping coefficient was $\alpha \approx 0.2$.[15]

To conduct micromagnetic simulations, we used the Object-Oriented MicroMagnetic Framework (OOMMF) software.[16] The sample volume was (120 × 120 × 1) μm$^3$ with the size of a single cell set to (1 × 1 × 1) μm$^3$. The uniaxial magnetic anisotropy field parameter was artificially increased to $K_u$ = -600 kJ/m$^3$ in order to keep the dynamics of the magnetization fixed within the sample's plane. This simplification effectively confined the problem to two dimensions, with two easy axes along [110] or [-1-10]. Additionally, the strong damping observed in the system posed a challenge during our simulation. With ultrashort excitation pulses (~1 ps), the simulation requires intolerably short time steps while producing oscillations that decay too rapidly. It is important to note that in the experiment, the level of damping experienced by the magnetization depends on the cone angle of precession, but such time-varying damping cannot yet be incorporated in micromagnetic simulations. To therefore overcome these difficulties, we used a realistic damping constant set to 0.2 but increased the pulse duration to 60 ps. In turn, this led to appropriately scaled properties of the considered system (saturation magnetization parameter $M_S$ = 20.4×10$^5$ A/m and the cubic magnetocrystalline anisotropy field strength $K_1$ = 500 kJ/m$^3$). This substantially-stronger saturation magnetization allowed us to maintain the properties of our garnet in the simulation despite the modification of the anisotropy constants. Moreover, we note that in simulation effective anisotropy cubic and uniaxial fields via the relation $H = 2K/M_S$ are about 490 mT and 588 mT, respectively. The alteration of one parameter, therefore, necessitates modification of the other to obtain similar material properties in terms of acting magnetic fields. The laser-induced strain was introduced with the extension developed by Yahagi et al.,[17] described in terms of a magneto-elastic field pulse with details given in Ref. 9. In short, we assume that the anharmonic interaction of the phonon modes shifts the in-plane coordinates with an amplitude proportional to the Gaussian profile of the pumping laser. This results in a macroscopic distribution of strains that is equivalent to the scenario of a non-uniformly heated object (e.g. a cylinder) with an axially-symmetric distribution of temperature. This gives rise to a magneto-elastic field containing spatial derivatives of the strain.[9]

Figure 1 presents the spatially-resolved results of the micromagnetic calculations, simulating the magnetization reversal driven by a strain pulse. The top panel contains images of the calculated magnetic patterns (see Fig. 1a,b) showing the evolution of magnetization at different times $\Delta t$ after the arrival of the strain pulse. To provide a point of comparison, the magneto-optical image in Fig. 1c shows the quadrupolar magnetic pattern that one obtains after experimentally exposing YIG:Co to a laser pulse of wavelength 14 μm.[9] At $\Delta t = 0$ – defined as the time at which the intensity of the strain pulse is maximum – the magnetization starts to precess in two opposite directions (black and white areas with the in-plane component of magnetization along [-110] and [1-10] directions in YIG:Co). After precession, at $\Delta t > 1$ ns, the magnetic domains are formed and are stable due to the non-zero coercive field of YIG:Co. The formation of two contrasting magnetic domain pairs is a result of laser-induced strain and the initial magnetization state.[9] The colored lines in the images are extracted from the cross-sections of the magnetization indicated in Fig. 1d. The solid black line shows the beginning of the domain-formation process. One can observe that the magnetization at the center of the irradiated area remains unchanged, while the spatial division of the magnetic pattern can already be discerned. Integration of half of the solid black line results in obtaining half of the Gaussian distribution (not shown here). The strongest strain appears in the part of the Gaussian beam's profile with the highest gradient, i.e., the steepest slope on its side. The red and blue lines in Fig. 1d are profiles of the switched magnetic domains obtained from simulation and experiment, respectively. The maxima of the black line fit very well within the experimentally-observed switched regions. In the middle of the experimental image (Fig. 1c) and its profile in Fig. 1d (blue line with circles), the demagnetization pattern can be seen, caused by the locally-excessive laser fluence. However, due to the low strain across the central part of the Gaussian beam, a uniform magnetic domain is not created here. The green line in Fig. 1d is the profile obtained at $\Delta t = 0$ but with the strength of the magneto-elastic field just below the switching threshold. A lower intensity results in the strain not being strong enough to create a domain. The signal's amplitude strongly depends on spatial localization, and the quadrupole symmetry is also visible. As can be seen again, the peaks, corresponding to the region with the strongest strain, are confined to the area within the switched magnetic domains.

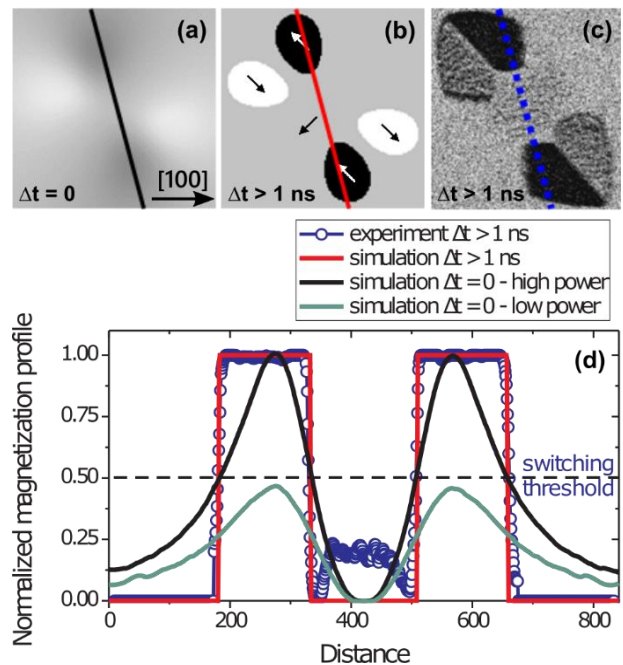

FIG. 1. Simulation of the switching regime. (a)-(b) Images of the calculated distribution of magnetization taken during and after the strain-induced switching; (c) Magneto-optical image of the magnetization in YIG:Co taken after excitation by a laser pulse of wavelength 14 μm. (d) Normalized profile of magnetization taken from the cross-sections marked in the images shown in panels (a)-(c). The red and blue profiles correspond to the switched domains from the simulation and experiment, respectively. The black line is a profile for switching power at the time of strain pulse maximum intensity $\Delta t = 0$. The green line was obtained at the same time $\Delta t = 0$ but with a magneto-elastic field

with strength below the switching threshold. The dashed black line marks the switching threshold. The arrows in (b) mark the magnetization directions.

In Fig. 2, we present the calculated results showing how the magnetization reversal depends on the amplitude of the magneto-elastic field pulse. Fig. 2a shows images of the magnetic pattern for different magneto-elastic field values ($H_{me1}$–$H_{me6}$). The red and blue colors correspond to magnetic domains with opposite magnetization, as marked with arrows (see also in Fig. 1b). The dashed line in the graph below marks the threshold level below which magnetic switching is not achieved. The size of the switched magnetic domains increases linearly with the magneto-elastic field (red line).[9] In the simulations, we do not observe any saturation of the switched area, i.e., the domain grows continuously in size with increasing field value (see Fig. 2b). In the experiment, in contrast, a continued increase of the laser fluence results in irrevocable damage to the sample due to heating.[9]

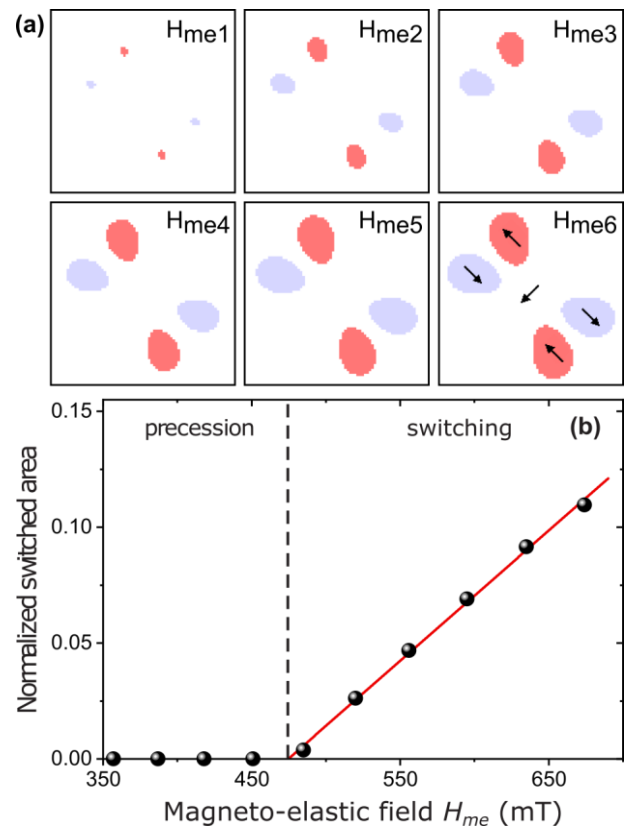

FIG. 2. (a) Calculated spatial distribution of the switched magnetization with increasing magneto-elastic field value ($H_{me1}$–$H_{me6}$ correspond to 485, 520, 557, 595, 635, and 674 mT, respectively) and (b) the magneto-elastic field dependence of the normalized switched area. The threshold power is marked with the black dashed line, while the red line is a linear fit. The normalized switched area is calculated as the ratio of the single switched domain area to the area of the field distribution. The arrows mark the magnetization directions.

After analyzing the switched magnetic pattern across equilibrium timescales, we proceed to study how strain affects the magnetization dynamics across sub-nanosecond timescales. In Fig. 3a, we present the stabilized distribution of magnetization following strain-induced excitation. The right and left panels were obtained with the magnetization initially uniformly oriented along the [110] and [-1-10] directions (marked +M and –M, respectively). Such a configuration resembles the visualization of a monodomain magnetic state in YIG:Co, achieved by applying an external magnetic field. We analyzed the calculations for three probe spot positions relative to the switched magnetization area for $H_{me} > H_{me1}$ (Fig. 3a). The calculated precession of magnetization at position "1", obtained for the opposite initial magnetization orientation with low magneto-elastic field value $< H_{me1}$ (i.e., below the switching threshold), is shown in Fig. 3b. Depending on the initial state, the magnetization vector starts to precess with different phase. Similarly, we observed a change of phase precession when probing at position "3" (not shown here), whereas we observe no change in magnetization at position "2". The spatial separation between the two oppositely-oriented magnetic domains (at positions "1" and "3") results in the simultaneous emergence of precession with opposite phases that exactly compensate for each other. When pumping with magneto-elastic field amplitude greater than the previously-identified threshold $> H_{me1}$ (see Fig. 2b), we observe a permanent switching of the $M_{[110]}$ component at probe positions "1" and "3" (blue and red lines in Fig. 3c, respectively). The precession after 20 ps is significantly damped since it is aligned mostly along the adjacent axis of anisotropy (see inset). There is no switching at position "2" due to the relation between the directions of the strain and the initial magnetization vector.

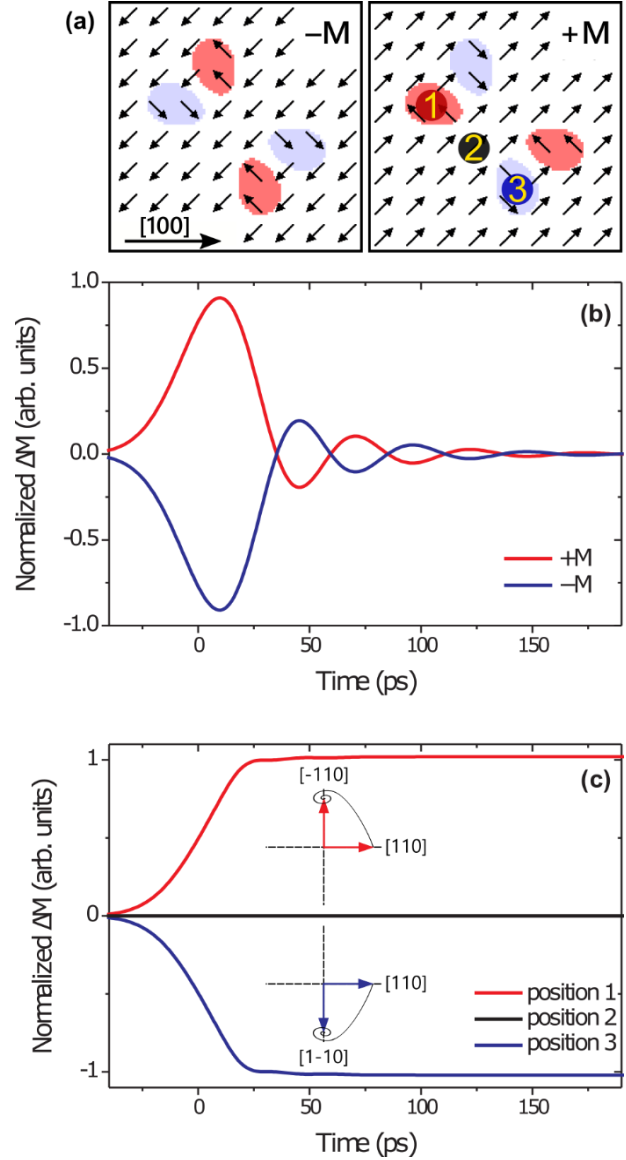

FIG. 3. Micromagnetic simulation of strain-induced magnetization dynamics. (a) The magnetization state after magneto-elastic field pulse for opposite initial states $M_{[110]}/M_{[-1-10]}$ (+M/–M) with different locations of the probe spot. (b) Precession of the magnetization component $M_{[-110]}$ for the two different starting magnetic states as indicated. (c) Magnetization switching for different spatial localization of the probing (positions "1"-"3") with initial magnetization +M.

Complimentary experimental measurements of the phonon-induced magnetization dynamics were obtained using the setup presented in Fig. 4a. The magnetization dynamics were measured using an ultrafast two-color pump-probe technique at the FELIX facility in the Netherlands.[18] In order to excite at resonance a longitudinal optical phonon mode in YIG:Co, we used a single 1-ps-long optical pulse with central wavelength λ = 14 μm delivered by the pulse-sliced free-electron laser.[19] The fluence of this pump pulse was set below the threshold required to achieve permanent switching of magnetization.[9] The probe was provided by a synchronized fiber-based oscillator (Orange High Power, Menlo), which delivers 150-fs-long optical pulses with a central wavelength of 520 nm. The linearly-polarized probe is delayed in time with respect to the pump by our use of a retroreflector mounted on a motorized translation stage. Upon transmitting through the YIG:Co sample, the probe's polarization is rotated due to the magneto-optical Faraday effect, providing sensitivity to the out-of-plane component of magnetization $M_Z$ at the focused probe's position. This polarization rotation is measured using a polarizing beam-splitter and two balanced photodetectors. By changing the spatial position of the tightly-focused probe pulse in relation to that of the pump pulse, we are able to measure magnetization dynamics across different parts of the quadrupolar domain pattern. The position of the probe pulse relative to the quadrupolar magnetic domains was obtained using *in-situ* static magneto-optical microscopy (similar to that used in Ref. 9). Our measurements were performed in the presence of a constant external magnetic field applied at a direction tilted about 18° away from the sample's [110] crystal axis, with amplitude $H = \pm15$ mT.

Figure 4b presents the magnetization dynamics obtained for two opposite field directions at the probe's position "1" (analogous to that shown in Fig. 3a). Depending on the initial in-plane direction of magnetization, the opposite phase of the magnetization precession is observed. Similar behavior is seen for simulations with different initial magnetization directions (see Fig. 3b), which confirms the magnetic nature of the observed signal. The step-like precession observed experimentally across the first sixty picoseconds may be assigned to the strain-induced effect of magneto-elastic energy modification, while the induced magnetic anisotropy has a characteristic rise-time of ~20-30 ps.[20] We fitted the experimental data with a damped sine function (solid lines),[15] omitting the initial step, showing that the magnetization precession has a frequency of 2.34 GHz. This frequency is similar for laser-induced photo-magnetic precession in YIG:Co with the laser pump wavelength of 1.3 μm.[21] The period of the magnetization precession observed experimentally is much longer than that seen in the micromagnetic simulations. However, we must emphasize that the numerical calculations were performed for the in-plane orientation of magnetization due to the strong "easy-plane" type of anisotropy. The period of magnetization precession can be tuned by modifying the values of the magnetization saturation $M_S$ and anisotropy constants $K_1$ and $K_u$. It is important to note that such an operation influences also the switched magnetization pattern, necessitating proper balancing of these material parameters as well as the strength and duration of the pulsed magneto-elastic field. We also studied experimentally the dynamics for three different positions of the probe beam (according to Fig. 3a), with the results shown in Figure 4c. Depending on the position, the signal is positive, negative, or in-between, but the phase remains unchanged throughout. Our detection of the out-of-plane component of magnetization does not provide information about the direction of the precession along [110] and [1-10] axes of YIG:Co. Due to that, the phase does not differ for the different probe locations. However, the measurements obtained at positions "1" and "3" also contain the step-like offset with opposite phases. This result suggests a different direction of the magneto-elastic anisotropy term.[9]

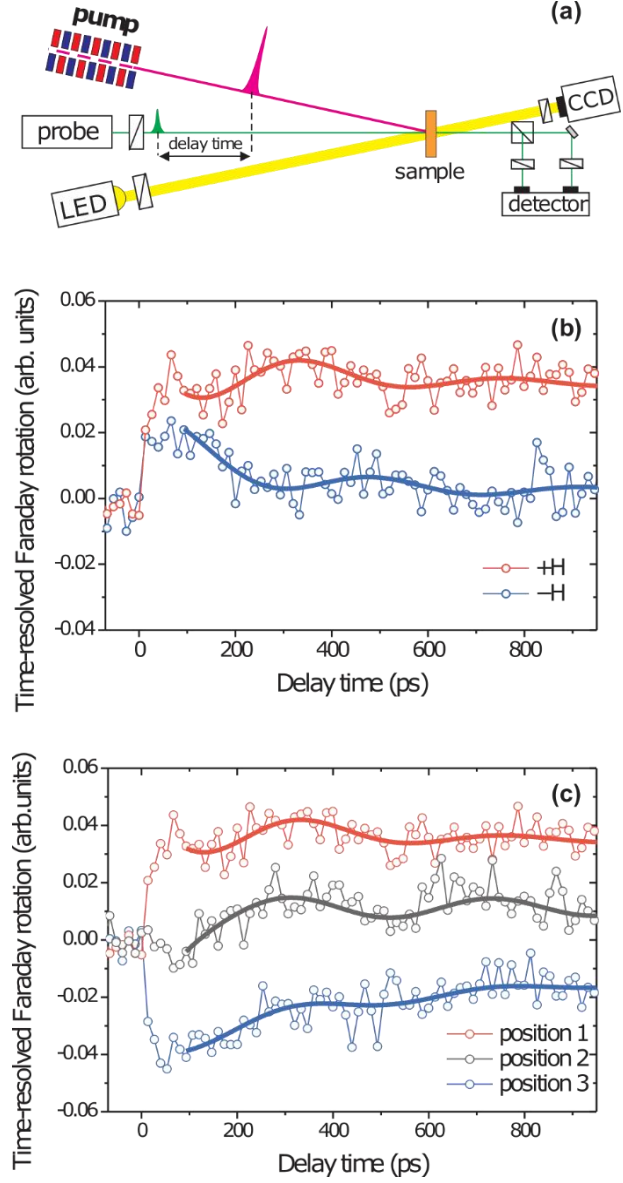

FIG. 4. The time-resolved Faraday rotation with a single infrared pump pulse of wavelength 14 μm. (a) Scheme of the experimental setup. (b) Magnetization precession measured with opposite orientations of the external magnetic field *H* and (c) precession measured by focusing the probe pulse at three different positions as indicated (see Fig. 3a). The solid colored lines in panels (b) and (c) are damped sine function fittings.

In conclusion, we have used micromagnetic simulations and experimental pump-probe measurements to study magnetization dynamics driven by strongly-coupled optical phonons that are excited at resonance. This excitation triggers a displacement of the equilibrium atomic positions, which is equivalent to macroscopic crystallographic strains. The obtained results reveal a strong dependence of the magnetization precession on both the probe's position and the external magnetic field. The initial step-like offset confirms that the precession mechanism arises from the strain-induced magnetic anisotropy through magneto-elastic energy term. The results presented here reveal the dynamical process of phononic manipulation of magnetization, suggesting that the process of phononic switching takes place across a timescale shorter than 100 ps.

Work was supported by the Foundation for Polish Science (POIR.04.04.00-00-413C/17-00). The authors thank all technical staff at FELIX for providing technical support.

The authors have no conflicts to disclose.

DATA AVAILABILITY. The data that support the findings of this study are available from the corresponding author upon reasonable request.